\documentclass[runningheads]{llncs}
\usepackage[T1]{fontenc}
\usepackage{graphicx}
\usepackage{amsmath}  
\usepackage{amssymb}  
\usepackage{booktabs}
\usepackage{multirow}
\usepackage{hyperref}

\def\B#1{\mathbf #1}
\def\C#1{\mathcal #1}
\def\R#1{\mathbb #1}

\begin{document}
%
\title{Multi-modal Food Recommendation using Clustering and Self-supervised Learning}
\titlerunning{Multi-modal Food Recommendation}

\author{Yixin Zhang\inst{1,2} \and
Xin Zhou\inst{3} \and
Qianwen Meng\inst{1,2} \and
Fanglin Zhu\inst{1,2} \and \\
Yonghui Xu\inst{2} \and
Zhiqi Shen\inst{4,5} \and
Lizhen Cui\inst{1,2}}

\authorrunning{Y. Zhang et al.}

\institute{School of Software, Shandong University, China \and
Joint SDU-NTU Centre for Artificial Intelligence Research (C-FAIR), \\ Shandong University, China \and
Alibaba-NTU Global e-Sustainability CorpLab (ANGEL), \\ Nanyang Technological University, Singapore \and
Joint NTU-UBC Research Centre of Excellence in Active Living for the Elderly, \\ Nanyang Technological University, Singapore \and
College of Computing and Data Science, Nanyang Technological University, Singapore
\\
\email{\{yixinzhang,mqw\_sdu,zfl\}@mail.sdu.edu.cn, \{xin.zhou,zqshen\}@ntu.edu.sg, xu.yonghui@hotmail.com, clz@sdu.edu.cn}}

%
%

%
\maketitle              
\begin{abstract}
Food recommendation systems serve as pivotal components in the realm of digital lifestyle services, designed to assist users in discovering recipes and food items that resonate with their unique dietary predilections. Typically, multi-modal descriptions offer an exhaustive profile for each recipe, thereby ensuring recommendations that are both personalized and accurate.
Our preliminary investigation of two datasets indicates that pre-trained multi-modal dense representations might precipitate a deterioration in performance compared to ID features when encapsulating interactive relationships. This observation implies that ID features possess a relative superiority in modeling interactive collaborative signals.
Consequently, contemporary cutting-edge methodologies augment ID features with multi-modal information as supplementary features, overlooking the latent semantic relations between recipes.
To rectify this, we present CLUSSL, a novel food recommendation framework that employs \underline{CLU}stering and \underline{S}elf-\underline{S}upervised \underline{L}earning. Specifically, CLUSSL formulates a modality-specific graph tailored to each modality with discrete/continuous features, thereby transforming semantic features into structural representation. Furthermore, CLUSSL procures recipe representations pertinent to different modalities via graph convolutional operations.
A self-supervised learning objective is proposed to foster independence between recipe representations derived from different unimodal graphs. Comprehensive experiments on real-world datasets substantiate that CLUSSL consistently surpasses state-of-the-art recommendation benchmarks in performance.

\end{abstract}
\section{INTRODUCTION}
Dietary intake is a critical determinant of human health and well-being, impacting both physiological and psychological states. Food recommendation systems (FRSs) aim to leverage user interaction data, multi-modal content, and ingredient information to personalize recipe suggestions that align with individual dietary needs. 
FRSs have seen notable advancements in recent years, driven by increased research interest and a wider range of lifestyle applications~\cite{rostami2023novel,recsys_workshop}.

With the pervasive presence of multi-modal data linked to recipes, contemporary advancements have witnessed the assimilation of abundant metadata, encompassing elements such as ingredients and visual attributes, to enhance the precision of recommendations.
For example, HAFR~\cite{HAFR} jointly models interaction data, ingredients, and visual features. Similarly, SCHGN~\cite{SCHGN} further considers higher-order relationships and calorie intake preferences. However, these methods predominantly leverage user, recipe, and ingredient IDs to capture collaborative signals, relegating multi-modal information to secondary feature roles. 
Meanwhile, our preliminary investigation (Table~\ref{intro_result}) demonstrates that directly utilizing continuous representations (embeddings) derived from multi-modal features (e.g., image and text embeddings) for collaborative filtering (CF) tasks may degrade recommendation performance. In contrast, using discrete features such as IDs yields superior results. Interestingly, employing ingredient information, also a discrete feature, as recipe embeddings leads to competitive performance. These findings suggest that discrete features are more effective in capturing the underlying structure of the data. Consequently, multi-modal models like FREEDOM~\cite{FREEDOM} propose an item-item graph construction based on similarities between transformed multi-modal features, where item IDs encapsulate the structural relationships derived from these features. 
While it can represent some level of similarity through item-item graphs, it fails to explicitly model the underlying categorical taxonomy (e.g., desserts, casseroles) that defines a crucial aspect of recipe semantics. This deficiency hinders its capacity to effectively integrate and discriminate between latent thematic categories within the culinary domain.
\begin{table}[t]
\caption{Performance of different modalities as item embedding w.r.t. LightGCN~\cite{LightGCN}.}
\def\arraystretch{1.0}%
\centering{
\begin{tabular}{l|cccc}
\toprule
\multicolumn{1}{c|}{\multirow{2}{*}{LightGCN}} & \multicolumn{2}{c}{Allrecipes} & \multicolumn{2}{c}{Food.com} \\ 
\multicolumn{1}{c|}{}                        & Recall@20       & NDCG@20      & Recall@20      & NDCG@20     \\          \midrule
\textemdash ID            & 0.1147          & 0.0572       & 0.1365         & 0.0960      \\   
\textemdash Ingredient    & 0.1140          & 0.0578       & 0.1466         & 0.1034      \\
\textemdash Image         & 0.1111          & 0.0557       & 0.1377         & 0.0932      \\
\textemdash Text          & 0.1136          & 0.0562       & 0.1365         & 0.0932      \\ 
\bottomrule
\end{tabular}}
\footnotesize{\begin{flushleft}{ID means default setting of~\cite{LightGCN}. For ingredient, we lookup ingredient embedding table and aggregate the corresponding representations as item embeddings. For image and text, we use the pre-trained features as item embeddings via an additional projector. }\end{flushleft}}
\label{intro_result}
\vspace{-10pt}
\end{table}

To this end, this work proposes CLUSSL, a novel multi-modal food recommendation model that employs clustering and self-supervised learning. 
Specifically, CLUSSL utilizes a novel two-stage approach to exploit multi-modal recipe information for recommendation. In the first stage, unsupervised clustering is applied to unimodal data (e.g., image features, text embeddings) with pre-trained continuous representations. These clusters act as ``prototype nodes'', summarizing the key semantic features within each modality. Subsequently, for each modality, a modality-specific graph is constructed to capture the relationships between these prototype nodes. This graph construction leverages ID features to encode the inherent structure of the data. Graph convolutional networks are then employed to effectively propagate and aggregate these semantic relationships across the graphs, allowing CLUSSL to exploit the rich relational information embedded within each modality.
Furthermore, CLUSSL incorporates a distance correlation constraint within a self-supervised learning framework. This constraint ensures that the learned recipe representations from different modalities (e.g., image-based and text-based) maintain a degree of independence. By combining these techniques, CLUSSL leverages the strengths of both unimodal data and multi-modal relationships, leading to demonstrably improved recommendation accuracy compared to existing baselines, as confirmed by extensive evaluations on real-world datasets.

\section{RELATED WORK}
In this section, we review the most relevant existing methods in food recommendation and multi-modal recommendation.

\subsection{Food Recommendation} 
In response to the diversified needs for food, FRSs are committed to providing accurate and personalized recommendation services for users. 
Recently, leveraging recipe metadata, such as ingredients and images, within the general CF framework has proven effective in mitigating data sparsity challenges and refining user preference modeling. 
HAFR~\cite{HAFR} develops a hierarchical attention mechanism to incorporate the feature interaction between metadata for recommendation. 
Moving beyond traditional methods, FGCN~\cite{FGCN} explores leveraging graph neural networks to capture high-order relationships through graph propagation and aggregation.
Furthermore, SCHGN~\cite{SCHGN} captures the user’s preference on food calories with self-supervised heterogeneous graph network. 
In addition, RecipeRec~\cite{RecipeRec} constructs heterogeneous recipe graphs to model the structure. 
Some recent works ~\cite{HealthRec,GreenRec} accommodates health criteria into food recommender to promote sustainable lifestyles~\cite{sus_survey}. 
\subsection{Multi-modal Recommendation} 
Multi-modal recommendation systems (MMRSs) strive to enhance items' representation for better recommendation performance by incorporating features from different modalities, including textual content and visual content~\cite{MMRec_Survey,MSM4SR,WhitenRec,DWSRec,zhang2021behavioral}. Early studies, such as VBPR~\cite{VBPR}, incorporate multi-modal contents with items' ID embeddings to extend the general CF framework. 
Inspired by the great success of graph-based recommendation methods~\cite{LightGCN,SelfCF,LayerGCN,GDCL,GCL4SR,zhao2024causal}, MMGCN~\cite{MMGCN} attempts to inject high-order semantics into user/item representations via message passing on modality-specific graphs. 
From another perspective, FREEDOM~\cite{FREEDOM} constructs item-item semantic graph from pre-trained multi-modal features to supplement the user-item interactions. 
An emerging trend is the integration of self-supervised learning methods into MMRSs, showcasing a notable boost. BM3~\cite{BM3} and LGMRec~\cite{LGMRec} improve user and item representation learning by adopting contrastive learning objective. DRAGON~\cite{DRAGON} utilizes homogeneous graphs based on item multi-modal information and a heterogeneous  bipartite user-item graph for effective recommendation. DGVAE~\cite{DGVAE} disentangles the user-item interactions and multi-modal information with graph learning.

\section{PRELIMINARIES}
In this section, we formulate the problem of food recommendation with multi-modal information and introduce the construction of the modality-specific graph. 

\subsection{Problem Formulation} 
We formally define the problem as follows. Let $\C{U}$ denote the set of users and $\C{I}$ denote the set of food recipes. The user-recipe binary interaction matrix is denoted by $\C{Y} \in \{0, 1\}^{|\C{U}| \times |\C{I}|}$, where $|\C{U}|$ and $|\C{I}|$ represent the cardinalities of users and recipes, respectively. Each entry $\C{Y}_{u, i} = 1$ indicates that user $u$ has interacted with recipe $i$. 
Associated with each recipe $i$ is the following modality information:
\romannumeral1) modality information with discrete features. We only consider recipe-ingredient relationships, denoted as $\C{M}_{\C{A}_i} \in \{0, 1\}^{|\C{A}|}$. The collection of ingredients across all recipes forms the complete ingredient set $\C{A}$, with cardinality $|\C{A}|$.
\romannumeral2) modality information with continuous features. In this work, we focus on visual features $\C{M}_{v_i} \in \R{R}^{d_v}$ and textual features $\C{M}_{t_i} \in \R{R}^{d_t}$. Specifically, $\C{M}_{v_i}$ are obtained from image of recipe $i$ through ResNet~\cite{ResNet}, while $\C{M}_{t_i}$ are extracted from textual description of recipe $i$ using a pre-trained T-5 model~\cite{T5}.

Based on the given notation, our objective is to predict the probability that user $u$ will consume recipe $i$, taking into account user-recipe interactions and the recipe's multi-modal information. 
\subsection{Construction of Modality-Specific Graphs}
Although heterogeneous graphs can include multiple types of vertices and edges, which helps to alleviate data sparsity and improve recommendation effectiveness~\cite{SCHGN,RecipeRec}, their complex structure poses significant challenges for model training. In this work, we propose constructing modality-specific bipartite graphs based on the discrete or continuous features for each modality. 
\subsubsection{Construct Bipartite Graph via Discrete Features.} Given the ingredient set $\C{A}= \{a\}$ and the recipe set $\C{I}= \{i\}$, the observed relationships between these sets are captured by the matrix $\C{R}^{\C{A}} \in \{0, 1\}^{|\C{A}| \times |\C{I}|}$, where each entry $\C{R}_{a,i}^{\C{A}} = 1$ if ingredient $a$ belongs to item $i$, otherwise $\C{R}_{a,i}^{\C{A}} = 0$. The $i$-th column denotes the discrete features $\C{M}_{\C{A}_i}$ of recipe $i$ regarding the ingredient modality. Based on the relationship matrix $\C{R}^{\C{A}}$, we construct a bipartite graph $\C{G}_{\C{A}}=\{(a, \C{R}_{a,i}^{\C{A}}, i)|a \in \C{A}, i \in \C{I}, \C{R}_{a,i}^{\C{A}}=1)\}$. 
\begin{figure*}
\centering
    \includegraphics[width=1.0\textwidth]{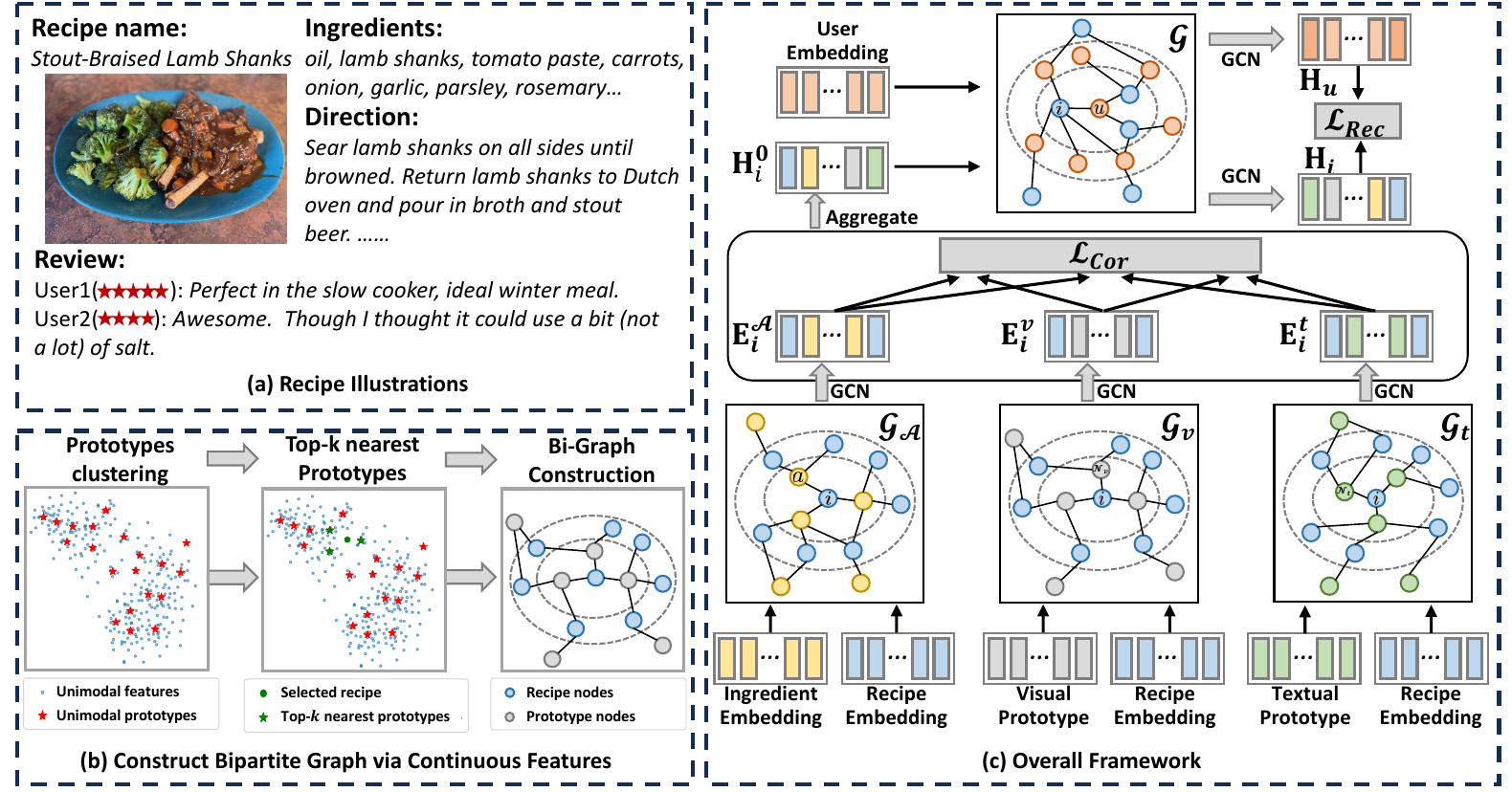}
    \caption{(a) exemplifies a recipe with its associated modalities, (b) showcases the construction of the modality-specific graph via continuous multi-modal features, and (c) illustrates  the overall framework of the proposed CLUSSL.}
    \label{fig:overall}
    \vspace{-20pt}
\end{figure*}
\subsubsection{Construct Bipartite Graph via Continuous Features.} According to~\cite{proto1}, closely binding item representations with raw modality features learned by pre-trained model can be detrimental. Thus, instead of using raw feature as an individual node representations, we introduce modality-specific cluster centres as ``prototype nodes''. Specifically, taking visual features as an example, we utilize K-means to cluster the modality feature vectors into $|\C{N}_v|$ cluster centers, i.e., $\C{N}_v$ denotes visual prototype nodes. Then we select the top $k$ nearest prototypes to define the visual-specific relationships for each recipe and form the matrix $\C{R}^{v} \in \{0, 1\}^{|\C{N}_v| \times |\C{I}|}$, where each entry $\C{R}_{n,i}^{v} = 1$ if prototype node $n$ belongs to top $k$ set of recipe $i$, otherwise $\C{R}_{n,i}^{v} = 0$. Based on the relationship matrix $\C{R}^{v}$, we construct a bipartite graph $\C{G}_{v}=\{(n, \C{R}_{n,i}^v, i)|n \in \C{N}_v, i \in \C{I}, \C{R}_{n,i}^v=1)\}$. Similarly, we construct the textual-specific bipartite graph $\C{G}_{t}=\{(n, \C{R}_{n,i}^t, i)|n \in \C{N}_t, i \in \C{I}, \C{R}_{n,i}^t=1)\}$. 
\section{METHODOLOGY}
This section delves into the individual components of CLUSSL. The overall architecture of the proposed method is illustrated in Fig.~\ref{fig:overall}. 
\subsection{Graph Collaborative Filtering Backbone}
In general, GNN-based CF methods~\cite{LightGCN} produce informative representations for users and items based on the message propagation and aggregation scheme, which exploit higher-order connectivity in the user-item graph and achieve state-of-the-art performance for recommendation. Following~\cite{BM3,FREEDOM,LGMRec} , we employ an efficient and effective LightGCN~\cite{LightGCN} as backbone to encode the structure of the user-recipe bipartite graph. 
Formally, we denote this user-recipe graph as $\C{G} = \{(u, \C{Y}_{u,i}, i)|n \in \C{U}, i \in \C{I}, \C{Y}_{n,i}=1)\}$ based on user-recipe interaction matrix $\C{Y} \in \R{R}^{|\C{U}| \times |\C{I}|}$. We obtain the adjacency matrix of the user-item graph as follows: 
\begin{align}\label{A}
    \B{A}=\begin{pmatrix}
 0        \; &  \; \C{Y}\\
 \C{Y}^T  \; &  \;  0
\end{pmatrix}.
\end{align}
Given $l$-th layer embedding matrix $\B{H}^l$, the simplifed message passing process in LightGCN is defined as: 
\begin{align}\label{gnn1}
    \B{H}^{l+1} = (\B{D}^{-1/2}\B{A}\B{D}^{-1/2})\B{H}^l,
\end{align}
where $\B{H}^0 \in \R{R}^{(|\C{U}|+|\C{I}|) \times d}$ is the $0$-th layer embedding matrix. User embeddings are randomly initialized and recipes embeddings are learned through different modality-specific graphs (introduced in section~\ref{sec:CL}). $\B{D}$ is a $(|\C{U}| + |\C{I}|) \times (|\C{U}| + |\C{I}|)$ diagonal matrix, also called as degree matrix, and the node embeddings of the $(l+1)$-th layer are only linearly aggregated from the $l$-th layer with a symmetrically normalized matrix $\B{D}^{-1/2}\B{A}\B{D}^{-1/2}$. Lastly, representations from all hidden layers are aggregated through a readout function to obtain the final embedding matrix used for recommendation: 
\begin{align}\label{gnn2}
    \B{H}_u &= \text{READOUT}(\B{H}_u^0, \B{H}_u^1, \dots, \B{H}_u^L),\nonumber\\
    \B{H}_i &= \text{READOUT}(\B{H}_i^0, \B{H}_i^1, \dots, \B{H}_i^L),
\end{align}
where $\B{H}_u$ and $\B{H}_i$ denote the final representations of user $u$ and recipe $i$, respectively, the READOUT function can be any differentiable function. Common designs include weighted sum, last-layer only, and others. We use the default mean function~\cite{FREEDOM,BM3} in practice. 

To generate recipe recommendations for user $u$, we first predict the interaction scores between the user and candidate recipes. Then, we rank candidate recipes based on the predicted scores in descending order and select the top-\textit{k} recipes as recommendations for the user. The interaction score is calculated as:
\begin{align}\label{y}
     \hat{y}_{u,i} = {\B{H}_u}^T \B{H}_{i}, 
\end{align}
\noindent where $\hat{y}_{u,i}$ is the prediction score of user $u$ towards recipe $i$. A high score suggests that the user prefers the recipe. 

To capture the collaborative information from implicit feedback, we adopt Bayesian Personalized Ranking (BPR) loss~\cite{BPR} in model training. Specifically, BPR loss ensures that the prediction score of the observed interactions higher than sampled unobserved ones. Formally:
\begin{align}\label{lossrec}
     \C{L}_{Rec} = \sum_{(u,i^+,i^-) \in \C{O}} -\textit{log} \sigma (\hat{y}_{u,i^+}-\hat{y}_{u,i^-}),
\end{align}
where $\sigma$ is the sigmoid function, $\C{O} = \{(u,i^+,i^-)|\C{Y}_{u,i^+} = 1, \C{Y}_{u,i^-} = 0\}$ denotes the pairwise training data, and $i^-$ denotes a sampled recipe that user $u$ has not interacted with.

\subsection{Unimodal Graph Representation Learning} \label{sec:CL}
Given the modality-specific bipartite graphs $\C{G}_{\C{A}}$, $\C{G}_v$, and $\C{G}_t$ towards different modalities, we employ graph convolutional networks to perform information propagation and aggregation. Without losing generality, we use LightGCN as the graph encoder, which similar to Eq.~\ref{gnn1} and Eq.~\ref{gnn2} derived from the corresponding adjacency matrix Eq.~\ref{A}. We transform these discrete features into dense-valued vectors through embedding lookup tables: recipes ($\B{E}_\C{I} \in \R{R}^{|\C{I}| \times d}$), ingredients ($\B{E}_\C{A} \in \R{R}^{|\C{A}| \times d}$), image prototypes ($\B{E}_v \in \R{R}^{|\C{N}_v| \times d}$), and text prototypes ($\B{E}_t \in \R{R}^{|\C{N}_t| \times d}$), where $d$ is the embedding space dimensionality. Utilizing these embeddings as the initial representations ($0$-th layer), we perform graph propagation and aggregation on each modality-specific graph to obtain the final representations of recipe $i$, denoted as $\B{E}^\C{A}_i$, $\B{E}^v_i$, and $\B{E}^t_i$, respectively. 

Based on the embeddings mentioned above, we aggregate these embeddings to define the $0$-th embedding of recipe $i$ in Eq.~\ref{gnn1}:
\begin{align}\label{agg}
     \B{H}_i^0 = \text{Aggregate}(\B{E}^\C{A}_i, \B{E}^v_i, \B{E}^t_i).
\end{align}
In this work, we simply define $\text{Aggregate}(\cdot)$ as vector summation. The embedding $\B{H}_i^0$ encompasses holistic perspective of the recipe's multi-modal profile, which is further used to capture collaborative information from user-recipe interactions.

\subsection{Cross-modal Self-supervised Learning}
To ensure the stability of learning recipe representations across different modality-specific graphs, we propose a self-supervised regularization to encourage the recipe representations (i.e., $\B{E}^\C{A}_i$, $\B{E}^v_i$, and $\B{E}^t_i$) to preserve sufficient independent information and avoid information redundancy. Although we learn the representations from different modality-specific graphs, there might still be redundancy due to high-order aggregation. To further promote independence among these representations, we adopt the distance correlation~\cite{cor,eLiveRec} as a regularization technique.
Distance correlation can capture the 
relationship between representations learned from different modalities while encouraging them to retain informative independence.
Formally, we define it as follows: 
\begin{align}
    \mathcal{L}_{Cor} =\sum_{m,n \in \{\C{A}, v, t\}, m\neq n} \frac{dCov(\B{E}^m_i, \B{E}^n_i)}{\sqrt{dVar(\B{E}^m_i) \cdot dVar(\B{E}^n_i)}},
\end{align}
where $dCov(\cdot)$ is the distance covariance between two matrices, and $dVar(\cdot)$ represents its own distance variance. Distance correlation helps mitigate the representation collapse often induced by other self-supervised constraints by maximizing dependencies between different modality-specific graphs.

\subsection{Multi-task Learning}
The proposed CLUSSL can be optimized in an end-to-end manner through a unified objective function, defined as follows:
\begin{align}\label{loss}
     \C{L} = \C{L}_{Rec} + \lambda\sum_{i \in (i^+,i^-)}\C{L}_{Cor} + \eta \|\Theta\|_2^2, 
\end{align}
where $\lambda$ is trade-off coefficient that balance the contributions of the self-supervised loss. $\eta$ and $\Theta$ represent the $L_2$ regularization coefficient and model parameters, respectively. 

\section{EXPERIMENTS}
\subsection{Experimental Settings}
\subsubsection{Dataset.}
We conduct experiments on two datasets collected from real-world platforms \href{https://www.allrecipes.com/}{www.allrecipes.com} and \href{https://www.food.com/}{www.food.com}. Each recipe in the datasets includes its ingredients, texts, images and the corresponding ratings from users, as shown in Fig.~\ref{fig:overall}(a). Among them, the ratings in the range of [1,5], we treat each rating as an implicit feedback record. Following~\cite{HAFR,SCHGN}, we holdout the latest $30\%$ of interaction history to construct the test set, and split the remaining data into training $(60\%)$ and validation $(10\%)$ sets. The statistics of two datasets after preprocessing are summarized in Table~\ref{dataset}.
\begin{table}[h]
\centering
\caption{Statistics of the experimental dataset.}
\begin{tabular}{lll}
\toprule
Characteristics & Allrecipes & Food.com \\ \hline
\#Users         & 68,768     & 7,596    \\
\#Items         & 45,630     & 29,943   \\
\#Interactions  & 1,093,845  & 323,199  \\
\#Ingredients   & 19,987     & 5,098    \\
Sparsity (\%)   & 99.96      & 99.85    \\ 
\bottomrule
\end{tabular}
\label{dataset}
\end{table}
\subsubsection{Metrics and Evaluation.}
During training, we sample one negative recipe for each user-recipe pair in the training set. The performance of different methods is assessed by two widely used evaluation metrics: Recall@$K$ and Normalized Discounted Cumulative Gain@$K$ (denoted by NDCG@$K$), where K is empirically set to 10 and 20. For each metric, we first compute the accuracy for each user on the testing data, and then report the averaged accuracy for all testing users. Following~\cite{HAFR,SCHGN}, there are 500 sampled negative recipe with popularity bias for one user and her interacted recipes in the test set. 
\subsubsection{Baseline Methods.}
We compare the proposed CLUSSL with the following baseline methods.
\textbf{General Collaborative Filtering:} BPR~\cite{BPR}, NeuMF~\cite{NeuMF}, LightGCN~\cite{LightGCN}.
\textbf{Multi-Modal Recommendation:} VBPR~\cite{VBPR}, MMGCN~\cite{MMGCN}, BM3~\cite{BM3}. FREEDOM~\cite{FREEDOM}, LGMRec~\cite{LGMRec}. 
\textbf{Food Recommendation:} HAFR\footnote{https://github.com/elisagao122/HAFR}~\cite{HAFR}, FGCN~\cite{FGCN}, SCHGN\footnote{https://github.com/TAEYOUNG-SYG/SCHGN}~\cite{SCHGN}. 
\subsubsection{Implementation Details.}
All the baseline methods are implemented by PyTorch~\footnote{https://pytorch.org/} and evaluated on a NVIDIA TITAN RTX GPU card. For fair comparison, the hyper-parameters of baseline methods are selected following the original paper, and the optimal settings are determined based on the grid-search and validation set. Multi-modal baselines are implemented based on the unified MMRec framework~\cite{MMRec_Lib}. For the proposed CLUSSL, we empirically set batch size to 512, embedding size to 64, learning rates to 0.002 and 0.001 for Allrecipes and Food.com, respectively. Step decay of the learning rate is also adopted. The regularization coefficient $\eta$ is set to 0.01. The trade-off coefficient $\lambda$ tuned in \{0, 0.001, 0.005, 0.01, 0.05, 0.1, 0.5, 1\}. The number of GCN layers, prototype nodes are also searched in grad. The number of top-$k$ nearest prototypes are set to 6 and 10 for Allrecipes and Food.com, respectively. We repeat the experiments five times and report the average results.

\subsection{Performance Comparison}
The performance achieved by different recommendation methods on two real-world datasets are summarized in Table~\ref{performance}. From the table, we have the following observations.

\begin{table*}[t]
\def\arraystretch{1.2}
\caption{The recommendation performance achieved by different methods. The best results are in \textbf{boldface}, and the second best results are \underline{underlined}. }
\resizebox{\textwidth}{!}{
\begin{tabular}{l|cccc|cccc}
\toprule
\multirow{2}{*}{Method} & \multicolumn{4}{c|}{Allrecipes}                                & \multicolumn{4}{c}{Food.com}                                  \\ 
                        & Recall@10     & Recall@20      & NDCG@10       & NDCG@20       & Recall@10     & Recall@20     & NDCG@10       & NDCG@20       \\ 
\midrule
BPR                     & 0.0637 & 0.1101 & 0.0418 & 0.0555 & 0.0750 & 0.1255 & 0.0726 & 0.0877 \\
NeuMF                   & 0.0636 & 0.1106  & 0.0411 & 0.0550 & 0.0741 & 0.1253 & 0.0725 & 0.0879 \\
LightGCN                & 0.0668 & 0.1147  & 0.0430 & 0.0572 & 0.0829 & 0.1365 & 0.0802 & 0.0960 \\
\midrule
VBPR                    & 0.0667 & 0.1115  & 0.0453 & 0.0575 & 0.0757 & 0.1315 & 0.0757 & 0.0923 \\
MMGCN                   & 0.0669 & 0.1143  & 0.0440 & 0.0580 & 0.0767 & 0.1293 & 0.0744 & 0.0901 \\
BM3                     & 0.0676 & 0.1152  & 0.0442 & 0.0583 & 0.0810 & 0.1313 & 0.0793 & 0.0938 \\
FREEDOM                 & 0.0710 & 0.1191  & 0.0441 & 0.0585 & 0.0843 & 0.1405 & 0.0827 & 0.0996 \\
LGMRec                  & 0.0661 & 0.1119  & 0.0452 & 0.0586 & \underline{0.0849} & \underline{0.1409} & \underline{0.0847} & \underline{0.1008} \\
\midrule
HAFR                    & 0.0703 & 0.1142  & 0.0480 & 0.0608 & 0.0712 & 0.1248 & 0.0685 & 0.0855 \\
FGCN                    & 0.0666 & 0.1138  & 0.0426 & 0.0567 & 0.0756 & 0.1284 & 0.0725 & 0.0886 \\
SCHGN                   & \underline{0.0720} & \underline{0.1186}  & \underline{0.0480} & \underline{0.0617} & 0.0818 & 0.1338 & 0.0744 & 0.0906 \\
\midrule
CLUSSL               & \textbf{0.0880} & \textbf{0.1372}  & \textbf{0.0690} & \textbf{0.0822} & \textbf{0.0943} & \textbf{0.1516} & \textbf{0.1144} & \textbf{0.1262}             \\ 
\bottomrule
\end{tabular}}
\label{performance}
\end{table*}

\textbf{Firstly, the superiority of CLUSSL.} The proposed CLUSSL demonstrates significant superiority over general CF models, multi-modal recommendation models and the state-of-the-art food recommendation models across all dataset. This indicates that the proposed CLUSSL is exceptionally well-designed for food recommendation, effectively leveraging both multi-modal information with discrete and continuous features. 

\textbf{Secondly, the effectiveness of multi-modal features.} By incorporating multi-modal features reasonably, almost all models have achieved better performance. For instance, VBPR outperforms the backbone BPR. Multi-modal recommenders based on LightGCN (i.e., BM3, FREEDOM, and LGMRec) basically achieve better results than LightGCN alone. Additionally, HAFR and SCHGN take into account not only visual features but also and ingredients information, demonstrating superior performance over other multi-modal models on Allrecipes. 

\textbf{Thirdly, the variability of modal semantics on performance.} Almost all multi-modal models showcase improved recommendation results. However, we observe that the performance of BM3 is inferior to LightGCN on Food.com. Moreover, HAFR and SCHGN are designed to deeply integrate multi-modal features with the recipe embeddings for prediction, yet their performance on Food.com is less than satisfactory. One potential reason might be that interactive collaborative signals are more easily captured by deep networks in relatively dense datasets, leading complex multi-modal semantics to contribute adverse effects without well-designed modules. The proposed CLUSSL mitigates this issue by transforming semantic features into structural representation.

\textbf{Finally, the effectiveness of self-supervised learning.} SCHGN achieves better performance than HAFR by employing a self-supervised ingredient prediction objective. BM3 and LGMRec leverage contrastive learning as an auxiliary optimization function, achieving notable success. We also partially attribute significant improvements of CLUSSL to self-supervised learning. More in-depth discussions can be found in ablation study and sensitivity analysis. 

\subsection{Ablation Study}

To study the importance of each component of CLUSSL, we consider the following CLUSSL variants for evaluation: 
\romannumeral1) {w/o $\C{M}_{v_i}$}, 
\romannumeral2) {w/o $\C{M}_{t_i}$}, 
\romannumeral3) {w/o $\C{M}_{\C{A}_i}$}.
These variants involve removing specific modal features related to recipes, thereby excluding their respective modality-specific graphs. \romannumeral4) {w/o $\C{L}_{Cor}$}: we set $\lambda$ to 0 in Eq.~\ref{loss} to eliminate the self-supervised learning component. 
\begin{table}[]
\caption{The recommendation performance of CLUSSL and its variants.}
\centering{
\begin{tabular}{l|cccc}
\toprule
\multicolumn{1}{c|}{\multirow{2}{*}{Method}} & \multicolumn{2}{c}{Allrecipes} & \multicolumn{2}{c}{Food.com} \\ 
\multicolumn{1}{c|}{}                        & Recall@20       & NDCG@20      & Recall@20      & NDCG@20     \\          \midrule
w/o $\C{M}_{v_i}$   & 0.1214          & 0.0655       & 0.1477       & 0.1176                \\
w/o $\C{M}_{t_i}$   & 0.1240          & 0.0677       & 0.1479       & 0.1236                \\
w/o $\C{M}_{\C{A}_i}$  & 0.1268          & 0.0694       & 0.1477       & 0.1196                \\
w/o $\C{L}_{Cor}$                 & 0.1246          & 0.0621       & 0.1487       & 0.1110                \\

CLUSSL                             & 0.1372          & 0.0822       & 0.1516         & 0.1262      \\ 
\bottomrule
\end{tabular}}
\label{ablation}
\end{table}

Table~\ref{ablation} summarizes the performance of CLUSSL variants on Allrecipes and Food.com datasets. Several observations are noteworthy: CLUSSL consistently outperforms w/o $\C{M}_{v_i}$, w/o $\C{M}_{t_i}$, and w/o $\C{M}_{\C{A}_i}$ across both datasets, indicating that removing any multi-modal information leads to decreased performance. This underscores the utility of all modalities in enhancing recommendation accuracy.
Additionally, the effectiveness of these modalities varies. Generally, visual features prove more crucial than textual or ingredient features. Textual information often contains redundant data, while ingredient features are confined by predefined sets. In contrast, visual features vividly depict food appearance, color, shape, and composition, crucial for stimulating user interest and appetite, thereby offering significant advantages.
Furthermore, experiments w/o $\C{L}_{Cor}$ highlight the role of the self-supervised learning in improving recipe representations for food recommendation. By promoting distinctiveness across modalities, CLUSSL ensures independence among multi-modal information, thereby enhancing recommendation effectiveness.

\subsection{Parameter Sensitivity Study}

\begin{figure*}
\centering
\includegraphics[width=0.93\textwidth]{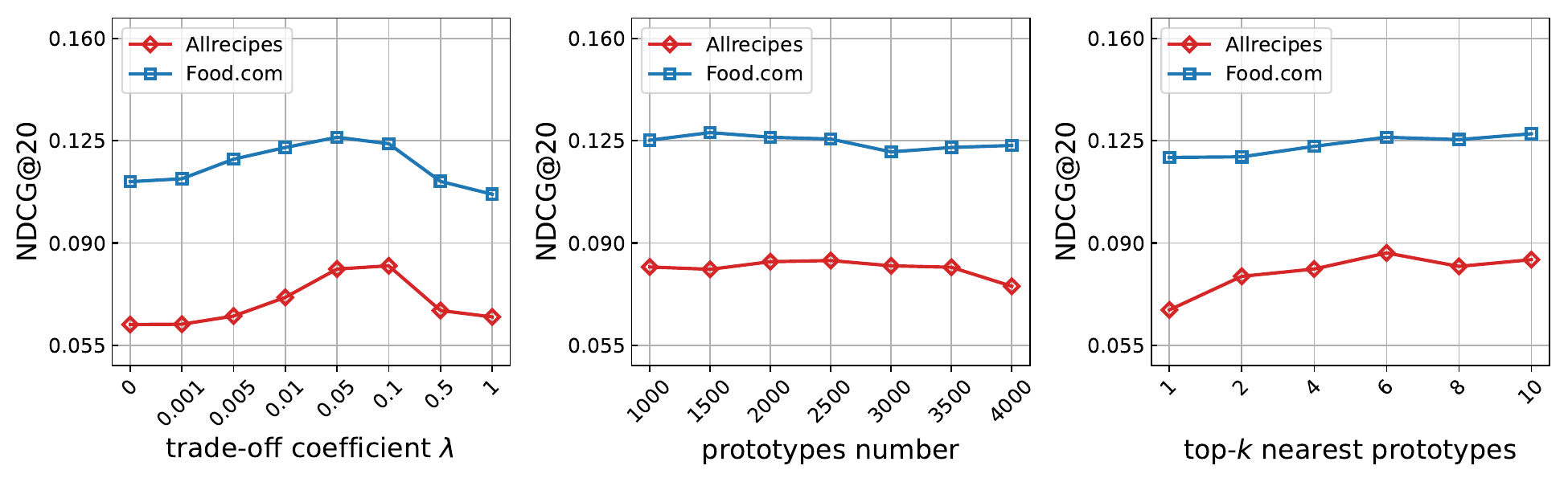}
    \caption{The performance trends of CLUSSL with respect to different settings of coefficient $\lambda$, prototypes number, and top-$k$ nearest on Allrecipes and Food.com datasets.}
    \label{fig:para}
\end{figure*}

We also perform experiments to study the impacts of three hyper-parameters: the trade-off coefficient $\lambda$, the number of prototypes in graph $\C{G}_{v}$ and $\C{G}_{t}$, and the top-$k$ nearest prototypes. Fig.~\ref{fig:para} shows the performance of CLUSSL with respect to different settings of these three hyper-parameters on both datasets. 

As depicted in Fig.~\ref{fig:para}, the results of the coefficient $\lambda$ on the two datasets showcase a consistent trend: the performance first improves to reach optimal and then declines as $\lambda$ increases. These results suggest that reasonable self-supervised constraints enable recipe representations to retain more comprehensive information, thereby enhancing recommendation performance. Regarding the number of prototypes, the performance trend appears relatively stable. Additionally, the best performance is achieved by setting top-$k$ to 6 and 10 on Allrecipes and Food.com datasets, respectively.

\section{CONCLUSION}
In this paper, we propose a novel food recommendation model, namely CLUSSL, which employs clustering to transform modal semantics into modality-specific graph across each modality. 
Moreover, CLUSSL leverages cross-modal self-supervised learning to encourage that recipe representations induced by different modality-specific graphs preserve sufficiently independent information. 
The backbone network can further benefit from independent representations with the self-supervised constraints, which providing superior performance. 
Experimental results on two real-world datasets verify that CLUSSL consistently outperforms state-of-the-art baseline recommendation models.

\subsubsection{Acknowledgements.}
This research is supported, in part, by the Natural Science Foundation of China (No.92367202, No.62202279), Outstanding Youth Science Fund Project (Overseas) of Shandong (No.2023HWYQ-039), Natural Science Foundation of Shandong Province (No.ZR2022QF018, No.ZR2023LZH006), 
the Youth Student Fundamental Research Funds of Shandong University, 
the State Scholarship Fund by the China Scholarship Council (CSC). 
This research is also supported, in part, by the RIE2025 Industry Alignment Fund – Industry Collaboration Projects (IAF-ICP) (Award I2301E0026), administered by A*STAR, as well as supported by Alibaba Group and NTU Singapore through Alibaba-NTU Global e-Sustainability CorpLab (ANGEL).

\bibliographystyle{splncs04}
\bibliography{clussl-arxiv}

\end{document}